\documentclass[sigconf, nonacm]{acmart}
\usepackage{caption}
\usepackage{subcaption}

\newcommand\vldbavailabilityurl{URL_TO_YOUR_ARTIFACTS}
\newcommand\vldbpagestyle{plain}

\definecolor{todocolor}{rgb}{0.2, 0.6, 1.0}

\begin{document}
\title{Stiff Circuit System Modeling via Transformer}

\author{Weiman Yan}
\affiliation{ University of Illinois at Urbana-Champaign
}
\email{weimany2@illinois.edu}

\author{Yi-Chia Chang}
\affiliation{University of Illinois at Urbana-Champaign
}
\email{yichia3@illinois.edu}

\author{Wanyu Zhao}
\affiliation{University of Illinois at Urbana-Champaign
}
\email{wanyu2@illinois.edu}

\begin{abstract}
Accurate and efficient circuit behavior modeling is a cornerstone of modern electronic design automation. Among different type of circuits, stiff circuit is challenging to model using previous frameworks. In this work, we propose a new approach using Crossformer, which is a current state-of-the-art Transformer model for time-series prediction tasks, combined with Kolmogorov–Arnold Networks (KANs), to model stiff circuit transient behavior. By leveraging the Crossformer’s temporal representation capabilities and the enhanced feature extraction of KANs, our method achieves improved fidelity in predicting circuit responses to a wide range of input conditions. Experimental evaluations on datasets generated through SPICE simulations of analog-to-digital converter (ADC) circuits demonstrate the effectiveness of our approach, with significant reductions in training time and error rates. 

\end{abstract}

\maketitle

\pagestyle{\vldbpagestyle}

\ifdefempty{\vldbavailabilityurl}{}{
\vspace{.3cm}
\begingroup\small\noindent\raggedright\textbf{Artifact Availability:}\\
The source code, data, and/or other artifacts have been made available at \url{https://github.com/yichiac/Crossformer/tree/circuit-var}.
\endgroup
}
\section{Introduction}

In the modern circuit design industry, the SPICE simulator is a cornerstone tool that enables accurate, early-stage validation of various designs. However, SPICE relies heavily on physics-based models and iterative numerical algorithms \cite{SPICE}, making it computationally expensive for large-scale circuit simulations. In addition, these models encode detailed design parameters such as transistor sizes and capacitances, which are highly confidential. When behavioral models are shared with downstream clients for integration and testing, such detailed information can potentially be reverse-engineered, posing risks to intellectual property (IP) protection. These limitations motivate the development of accurate black-box behavioral models that can approximate circuit responses without exposing internal design details.

Neural networks (NNs) have emerged as a promising alternative due to their strong approximation capability and flexibility. Prior work \cite{NN} applies NNs to circuit transient simulation, while \cite{RNN1,RNN2} further improves accuracy using recurrent neural networks (RNNs) and enables co-simulation with physics-based models. Although these approaches demonstrate improved practicality, they are primarily designed for discrete-time modeling. To address this limitation, \cite{CTRNN} leverages ODE-RNN \cite{RNNODE} to develop continuous-time circuit behavioral models, achieving state-of-the-art performance.

Despite these advances, modeling stiff circuit systems remains a significant challenge. Stiffness refers to systems in which certain components evolve slowly while others change rapidly. This phenomenon is common in circuit design, where high-frequency signals (e.g., clock edges) coexist with low-frequency analog dynamics. Existing ML-based models often struggle to capture such heterogeneous behavior, leading to degraded performance when applied to stiff systems. While prior works \cite{stiff1,stiff2} attempt to address stiffness using equation-based approaches, they are often difficult to implement and lack generality across different circuit types. Therefore, developing a robust black-box model that can effectively capture stiff dynamics remains an open problem.

From another perspective, stiff circuit behavior can be interpreted as inherently multi-modal, where different temporal components—such as fast switching transients and slow-varying signal envelopes—exhibit distinct characteristics. In this view, circuit responses are governed by multiple interacting modalities rather than a single homogeneous process. Such heterogeneity makes it difficult for conventional models to learn a unified representation. Recent advances in representation learning \cite{du2026unsupervised, liuzhipeng4, jasen4, jasen6} suggest that capturing interactions across different modalities, while preserving their unique characteristics, can significantly improve modeling fidelity. This perspective provides a useful lens for understanding the complexity of stiff circuit systems.

In this paper, we propose a new circuit behavioral model that is capable of capturing stiff dynamics across a wide range of input conditions. The model is designed to accurately predict circuit outputs under high-frequency inputs, low-frequency inputs, or mixtures of both, while maintaining flexibility across different circuit types.

To achieve this goal, we leverage attention-based architectures. With the emergence of Transformer models \cite{attention}, attention mechanisms have demonstrated strong performance in sequence modeling and time-series prediction. Prior work \cite{attentionODERNN} extends ODE-RNN with attention to improve time-series forecasting, while \cite{contiformer} combines Transformer architectures with Neural ODEs for continuous-time modeling. Motivated by these successes, we hypothesize that attention mechanisms are well-suited for capturing the multi-scale and multi-modal nature of stiff circuit dynamics. Specifically, we build upon the Transformer architectures proposed in \cite{contiformer,crossformer} and explore data representations tailored for both high-frequency and low-frequency signals.

\begin{figure*}[hpt]
    \begin{center}
        \includegraphics[width=0.9\textwidth]{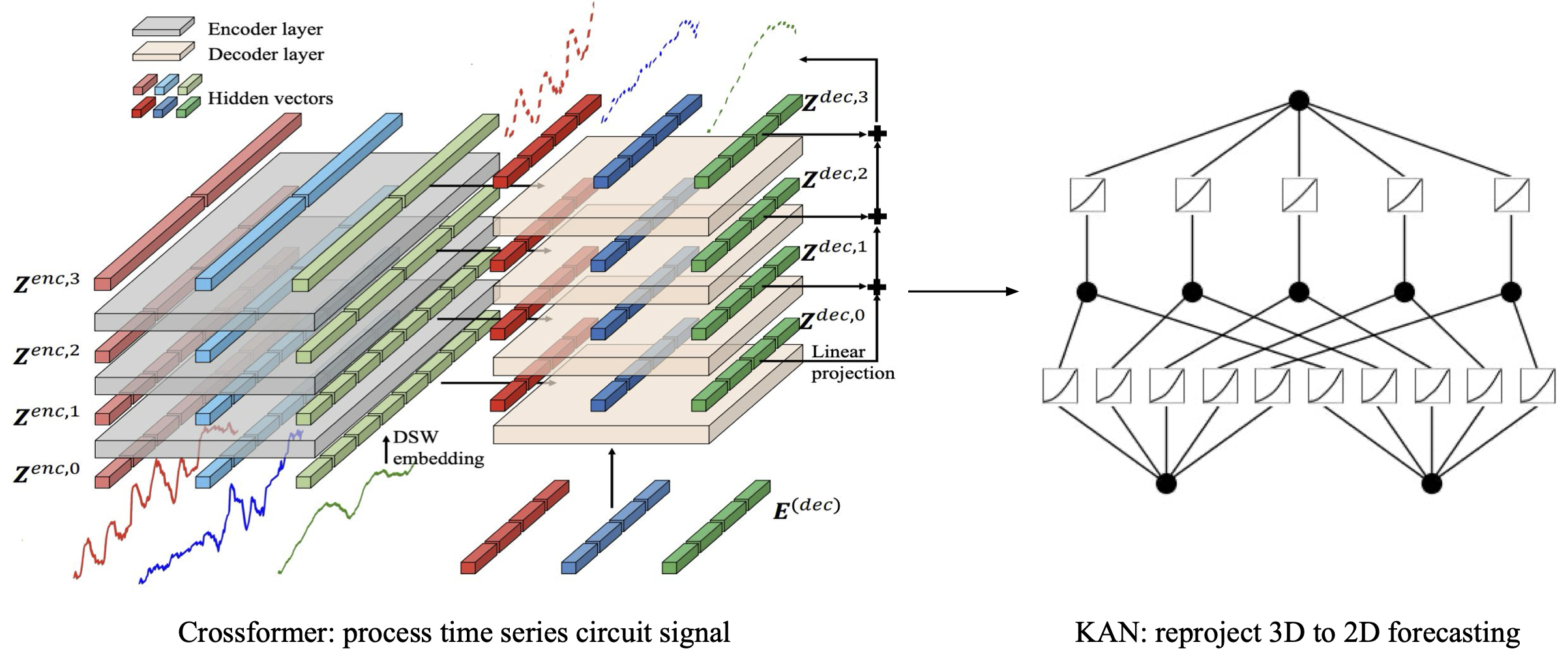}
    \end{center}
    \caption{Our model leverages Crossformer \cite{crossformer} and KAN \cite{KAN} for time-series forecasting. Specifically, Crossformer is used to incorporate input time-series information. Then, KAN is added to the decoder layer to project outputs to the desired forecasting dimensions.}
    \label{fig:architecture}
\end{figure*}

This work demonstrates the effectiveness of transformer-based architectures for modeling stiff circuit systems and highlights the potential of attention-based surrogate models in circuit design. By improving simulation efficiency and reducing computational cost, the proposed approach can significantly enhance circuit design workflows. Moreover, this work provides a new perspective on treating circuit dynamics as multi-modal temporal processes, which may inspire further research on machine learning-based surrogate modeling in the integrated circuit industry.

\section{Related works}
Previous work ~\cite{NN} uses Neural Networks (NNs) for circuit transient simulation. Chen et al. ~\cite{RNN2} and Luongvinh and Kwon ~\cite{RNN1} adopted Recurrent Neural Networks (RNNs) to further improve the accuracy of the learned behavioral model. In ~\cite{RNN2}, the trained model can be successfully integrated with other physical models to perform co-simulation, which makes the ML-based model more practical to be used in the industry. Nonetheless, the models proposed in ~\cite{NN,RNN1,RNN2} are all designed for discrete time data, but the input and output of a circuit in its nature should be continuous. This would introduce some errors when these models are used in practical conditions. 

Fortunately, Neural ODE ~\cite{NeuralODE} provides a way to make RNN continuous. ~\cite{RNNODE} introduces the ODE-RNN model by combining Neural ODE and Gated Recurrent Unit (GRU) ~\cite{GRU} to achieve state-of-the-art performance. ~\cite{CTRNN} developed a continuous time RNN model as an extension of ODE-RNN to be used in circuit behavioral modeling and reported a significant improvement in accuracy. 

One remaining problem in ~\cite{CTRNN} is that the model performance will be much worse when it is used to model a stiff circuit system. Stiffness refers to the situation when some parts of a system change very slowly, while other parts change very quickly. This is not a rare situation in circuit design. In fact, most of today’s circuits operate with input signals and output signals at very different frequencies, but unfortunately none of the previous ML-based circuit models can learn the stiff behavior effectively. Some previous works ~\cite{stiff1,stiff2} tried to address this problem via equation-based models, but they are hard to implement and not a universal solution to all types of circuits.  Furthermore, these equation-based models would add complexity to the simulation, which makes circuit system slower to simulate.

In recent years, with the emergence of Transformer models~\cite{attention}, attention mechanisms have demonstrated strong capability in improving performance on time-series tasks~\cite{liuzhipeng1, liuzhipeng2, liuzhipeng3}. For example,~\cite{attentionODERNN} incorporates attention into the ODE-RNN framework, showing that attention can enhance prediction accuracy in time-series modeling. Rather than extending recurrent architectures, ~\cite{contiformer} directly adopts the Transformer paradigm and extends it to the continuous-time domain, enabling more effective modeling of temporal dynamics. In addition, other approaches from the time-series community have explored alternative generative frameworks such as GANs~\cite{liuzhipeng5}. However, these methods have not been widely investigated in the context of circuit behavioral modeling.

\section{Methodology}

Given a circuit with $M$ inputs and $N$ outputs, we have the input time-series data $\boldsymbol{X}\in\mathbb{R}^{M\times T}$ and the output time-series data $\boldsymbol{Y}\in\mathbb{R}^{N\times T}$, where $T$ is the total number of time steps. The purpose of the circuit behavioral model is to predict the output matrix based on the input matrix.

As shown in Fig. \ref{fig:architecture}, we mainly leverage the structure of Crossformer ~\cite{crossformer}, which is the state-of-the-art model for multivariate time-series forecasting. Then, we add Kolmogorov-Arnold Networks (KANs) ~\cite{KAN} layers to the last layer of the Crossformer decoder to project the hidden vectors to the dimension of the circuit model.

\subsection{Crossformer} We chose Crossformer for two reasons. First of all, in each dimension, the Dimension-Segment-Wise (DSW) Embedding can divide the time series data into segments and embed them into the feature vector. Using segments instead of single data points enable the model to capture the rising edge and falling edge of the clock signal more easily. The rising and falling edges are usually the parts that cannot be accurately modeled by the previous framework. Secondly, the hierarchical encoder-decoder captures both cross-time and cross-dimension dependency, which aligns with the behavior of circuits.

\subsection{Kolmogorov-Arnold Networks} Recently, Kolmogorov-Arnold Networks (KANs) have emerged as alternative blocks to Multi-layer Perceptrons (MLPs) in deep learning models, demonstrating superior performance and interoperability. KANs use learnable activation functions instead of weights to model the input and output relationship. Since the learnable functions are represented by univariate functions, it gives KANs great advantages in AI for science jobs. Since most of circuits can be accurately represented by physics models and equations, we believe that KANs can provide better performance compared to MLPs for our task.
\subsection{Loss function}
Finally, the model is trained using Normalized Root Mean Square Error (NRMSE). The NRMSE can be written as

\begin{equation}
    NRMSE = \frac{1}{N\cdot T}\sum^N_{i=1}\sum^T_{j=1} (y^{(i)}_j - \hat{y}^{(i)}_j)^2,
\end{equation}
where $N$ is the number of output dimension, $T$ is the number of sample in each dimension, $y^{(i)}_j$ and $\hat{y}^{(i)}_j$ represent the ground truth and predicted voltage of the $i-$th output waveform at time index $j$. In our ADC test case, $N=2$ and $K=500$.

\section{Experimental Setup}

\subsection{Test Circuit}
\begin{figure}[h!]
\centerline{\includegraphics[width=\linewidth]{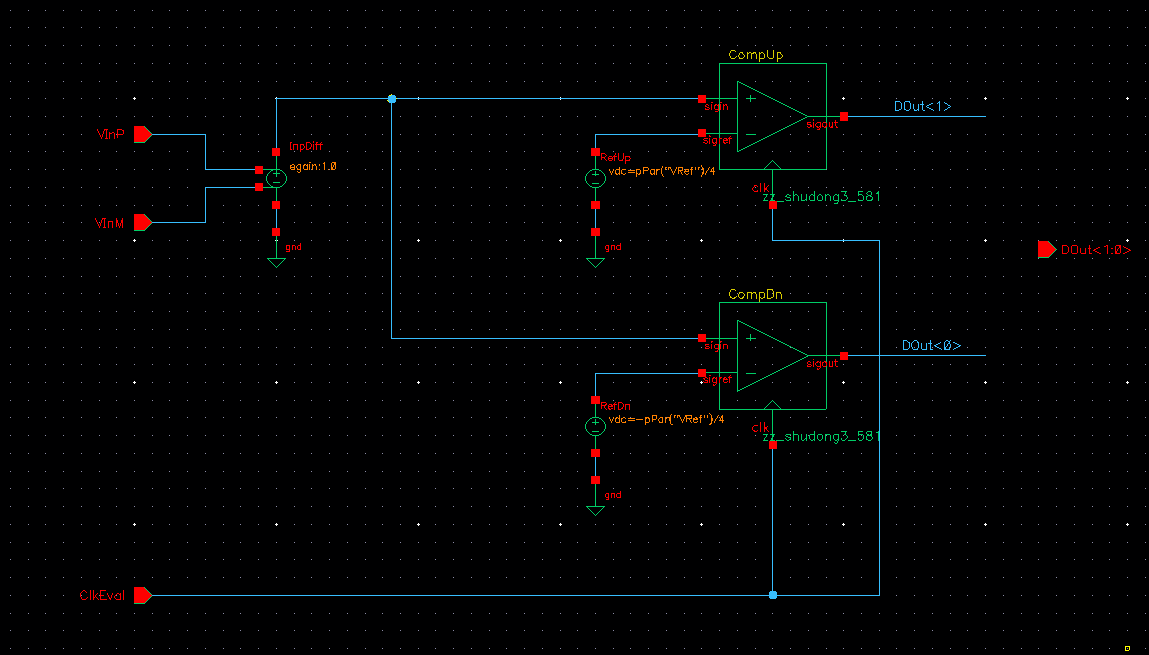}}
\caption{Schematic of our 1.5-Bit stage sub ADC.}
\label{ADC}
\end{figure}
To evaluate the performance of the models, we choose analog-to-digital-converter (ADC) circuit as the test case. In particular, our ADC is a 1.5-Bit stage sub ADC module (Fig. ~\ref{ADC}) that receives a pair of analog differential input (VInP and VInM) and a clock signal (ClkEval), then produce a pair of digital outputs (DOut<1> and DOut<0>). Under the expected operation condition, the clock signal is usually at high frequency, whereas the analog inputs and digital outputs are at relatively low frequency, which demonstrates the stiff behavior.

We implement the ADC using NCSU45 PDK, which is an open-source generic process design kit (PDK).

\subsection{Dataset Generation}

We use the SPICE simulator to generate a comprehensive dataset for training and evaluation. The dataset should cover the expected operation range of the ADC.

To produce real-world-like analog input signal, we use pseudorandom bit sequences (PRBS) data and feed it into a lossy channel implemented using a 175 Ghz RLC low pass filter. The channel will smooth the PRBS data waveform and remove the overly-high frequency part from the signal, which makes it more realistic. The output of the channel will be used as the analog inputs to the ADC circuit. In each data record, the bit time of the PRBS data will be a random number sampled uniformly from 50 ns to 150 ns. The rise-time and fall-time of the PRBS signal will also be a random number that uniformly ranges from 20\% to 30\% of the bit-time. The input PRBS data will have a constant common voltage at 0.45 V and a random differential mode voltage sampled uniformly within 0.15 V to 0.25 V.

For our clock signal in each data record, it will have a uniform random frequency ranging from 18.18 Mhz to 16.67 Mhz. The rise-time and fall-time of the clock signal will be fixed at 2500 ns for all the experiments. However, the clock will have random phase shift for each data record that ranges uniformly from $0^\circ$ to $180^\circ$. 

Besides, we also add randomness to the input and output resistance and capacitance for each data record to simulate different load conditions when the ADC is connect to other circuits. The resistance and capacitance will be uniformly sampled from 0$\Omega$ to 10$\Omega$ and 90pF to 110pF, respectively.

\begin{figure}[t]
\centerline{\includegraphics[width=\linewidth]{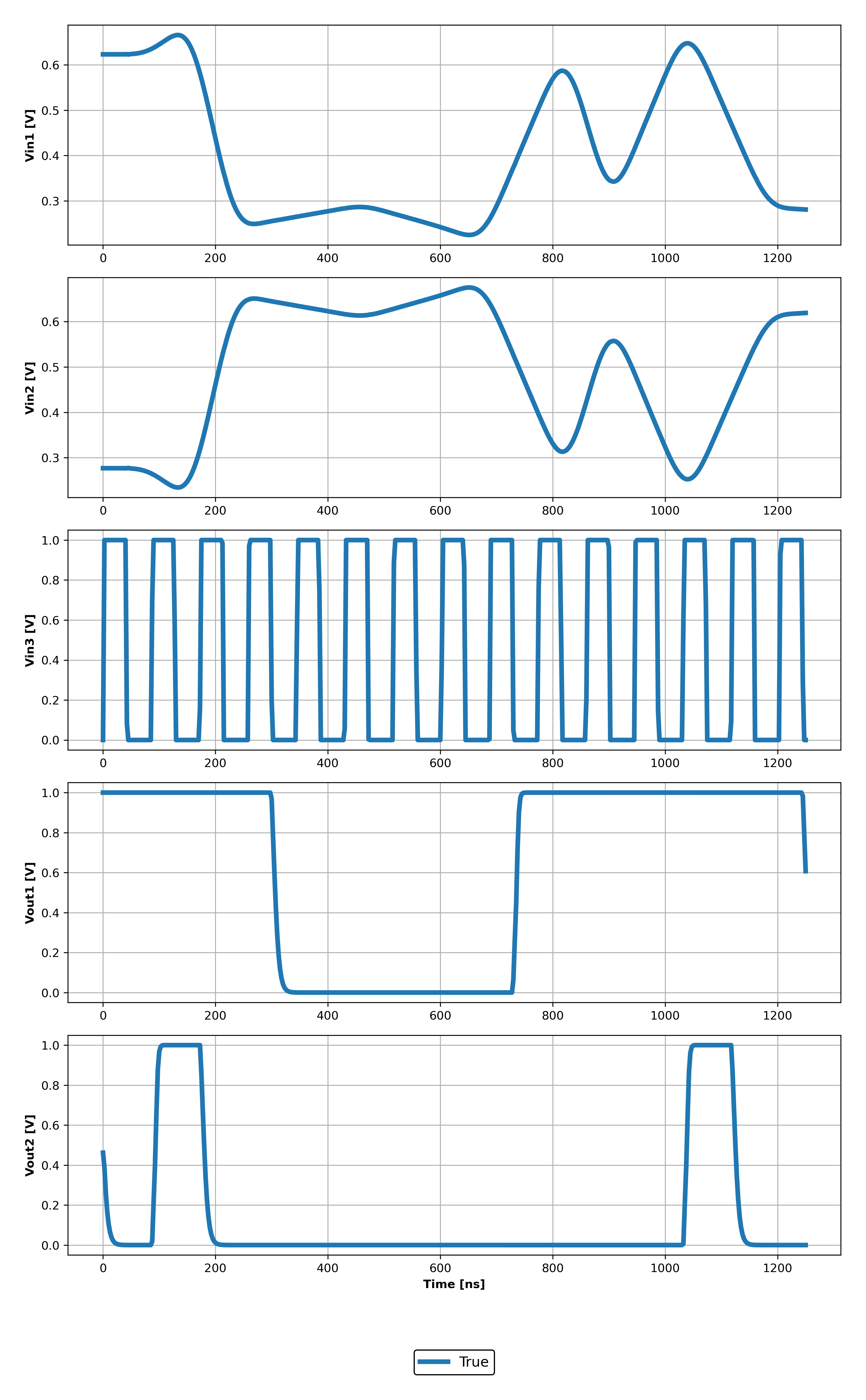}}
\caption{One example data record taken from our generated dataset. The top three columns are input signals with low and high frequencies. The bottom two columns are the ground truth outputs the model predictions want to fit. The stiffness in the output signals are challenging for time-series prediction models.}
\label{data_sample}
\end{figure}

To demonstrate the outlook of our dataset, we pick an arbitrary data record from it (Fig. ~\ref{data_sample}). In our dataset, each data record contains 5 rows. The top three rows are the input to our ADC circuit and the bottom two rows are the output waveform we obtained by performing transient simulation using SPICE. All the waveform have the same length, which is 1250 ns. We then sample the voltage of the waveforms every 2.5 ns to obtain the training data. This sample time is based on the  Therefore, each data record is a matrix with shape $5 \times 500$. Our dataset contains 2K such data records.

\subsection{Model Evaluation}
The dataset will be split into training (70\%), validation (15\%), and test (15\%) sets. We would use Normalized Root Mean Square Error (NRMSE) to measure the accuracy of the trained model on the validation dataset.

In addition, we also want to compare the size of dataset and time needed for model training. Since CTRNN is a sequential model which cannot be accelerated using GPU, it is unfair to directly compare our model training time with CTRNN. Therefore, we compare the number of epoch it takes for the models to converge. Note that CTRNN is trained using Cross Entropy (CE) loss since empirically it demonstrates better accuracy than NRMSE when we use CTRNN model.

For hyperparameter tuning in KAN, we experimented with different numbers of neurons: \{5, 10\} and grid intervals: \{5, 15, 50\}. To achieve the better training results, we also experimented with different learning rates: \{1e-3, 1e-4, 1e-5\}, optimizers: \{Adam, RMSProp\}, and early stopping criteria. For Crossformer, we experimented with hidden states dimensions (d\_model): \{256, 512\} to observe the effects of the larger hidden dimension.

\section{Results}
\begin{table}[h!]
\caption{Accuracy comparison between previous CTRNN, Crossformer only and our model. Crossformer + KAN can outperform CTRNN with lower NRMSE}
\centering
\begin{tabular}{|l|c|c|r|}
\hline
Model & Average NRMSE \\
\hline
CTRNN & 31.7\%\\
Crossformer only & 25.2\% \\
Crossformer + KAN (Our Model) & 21.1\% \\
\hline
\end{tabular}
\label{table1}
\end{table}

For experiment, we compared CTRNN and Crossformer with our proposed Transformer-based circuit models. The experiment result is summarized in Table ~\ref{table1}.

For the accuracy comparison shown in Table ~\ref{table1}, Crossformer is 20\% more accurate (from 31.7\% to 25.2\%) than the state-of-the-art model CTRNN, which verifies our thesis that the attention mechanism and Transformer architecture can better express the behavior of a stiff circuit. Our model, which combines Crossformer and KANs, further reduces the NRMSE from 25.2\% to 21.2\%.

\begin{figure}
     \centering
     \begin{subfigure}[b]{0.45\textwidth}
         \centering
         \includegraphics[width=\textwidth]{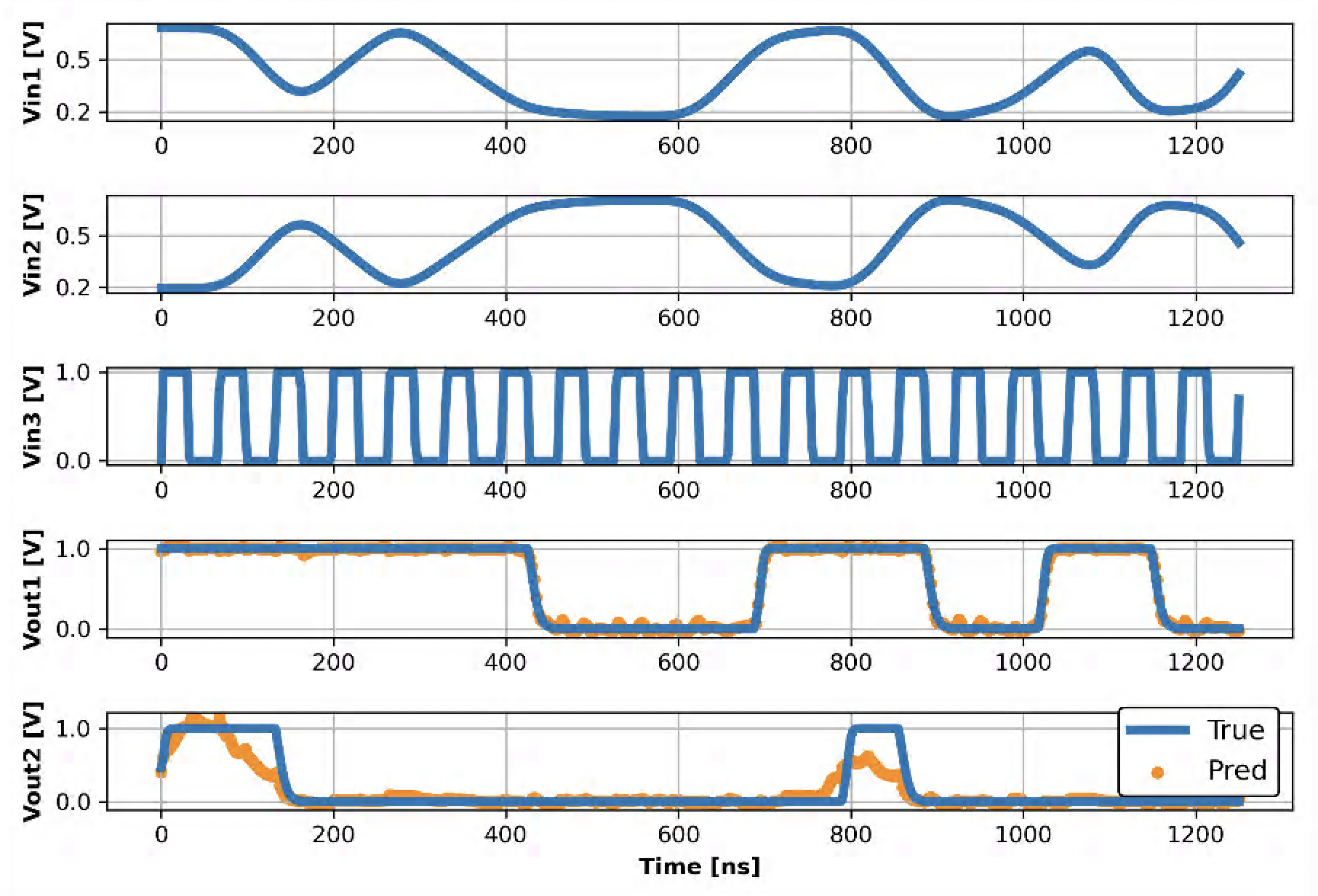}
         \caption{CTRNN Prediction}
     \end{subfigure}
     \begin{subfigure}[b]{0.45\textwidth}
         \centering
         \includegraphics[width=\textwidth]{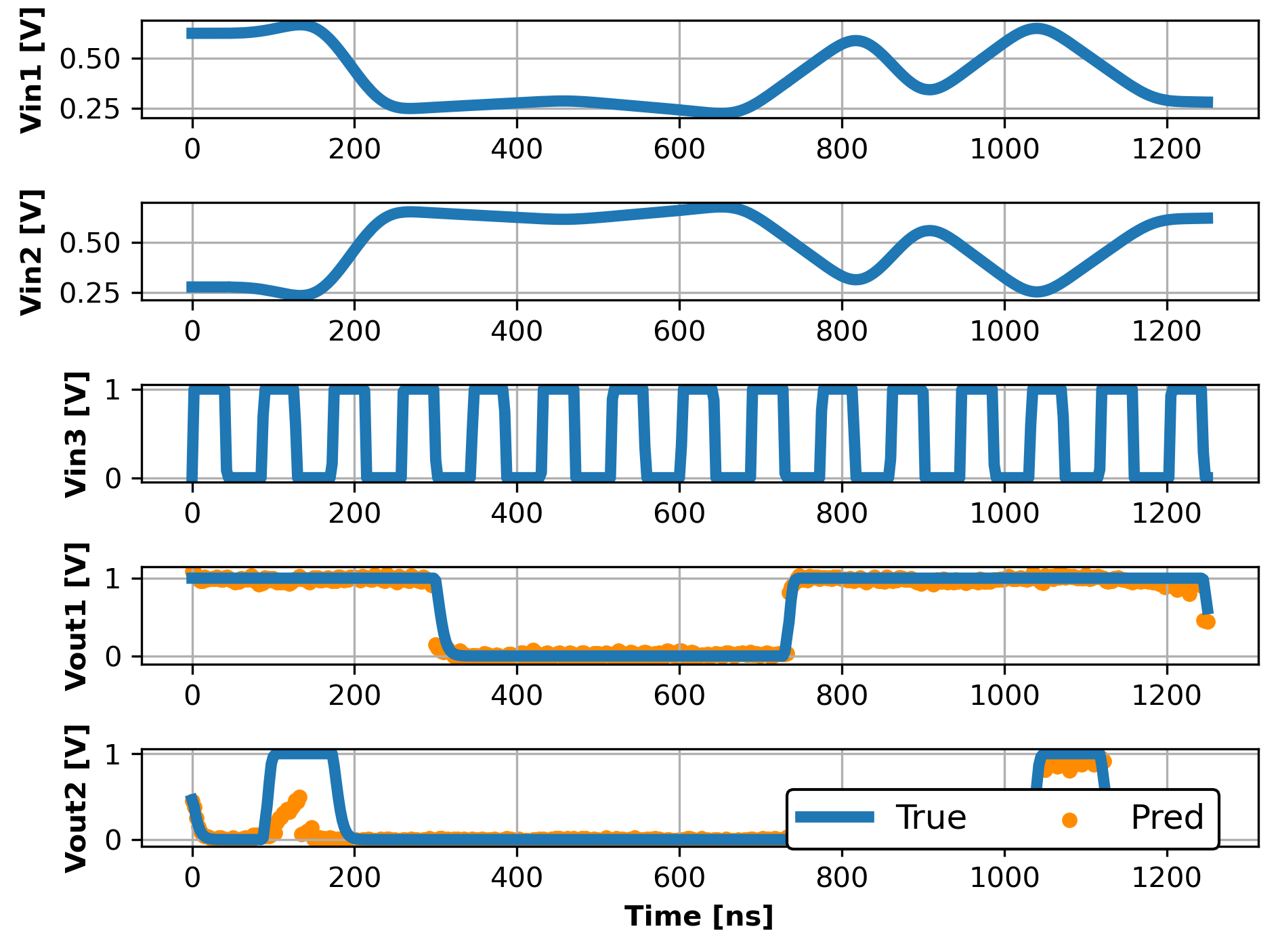}
         \caption{Crossformer Only Prediction}
     \end{subfigure}
     \begin{subfigure}[b]{0.45\textwidth}
         \centering
         \includegraphics[width=\textwidth]{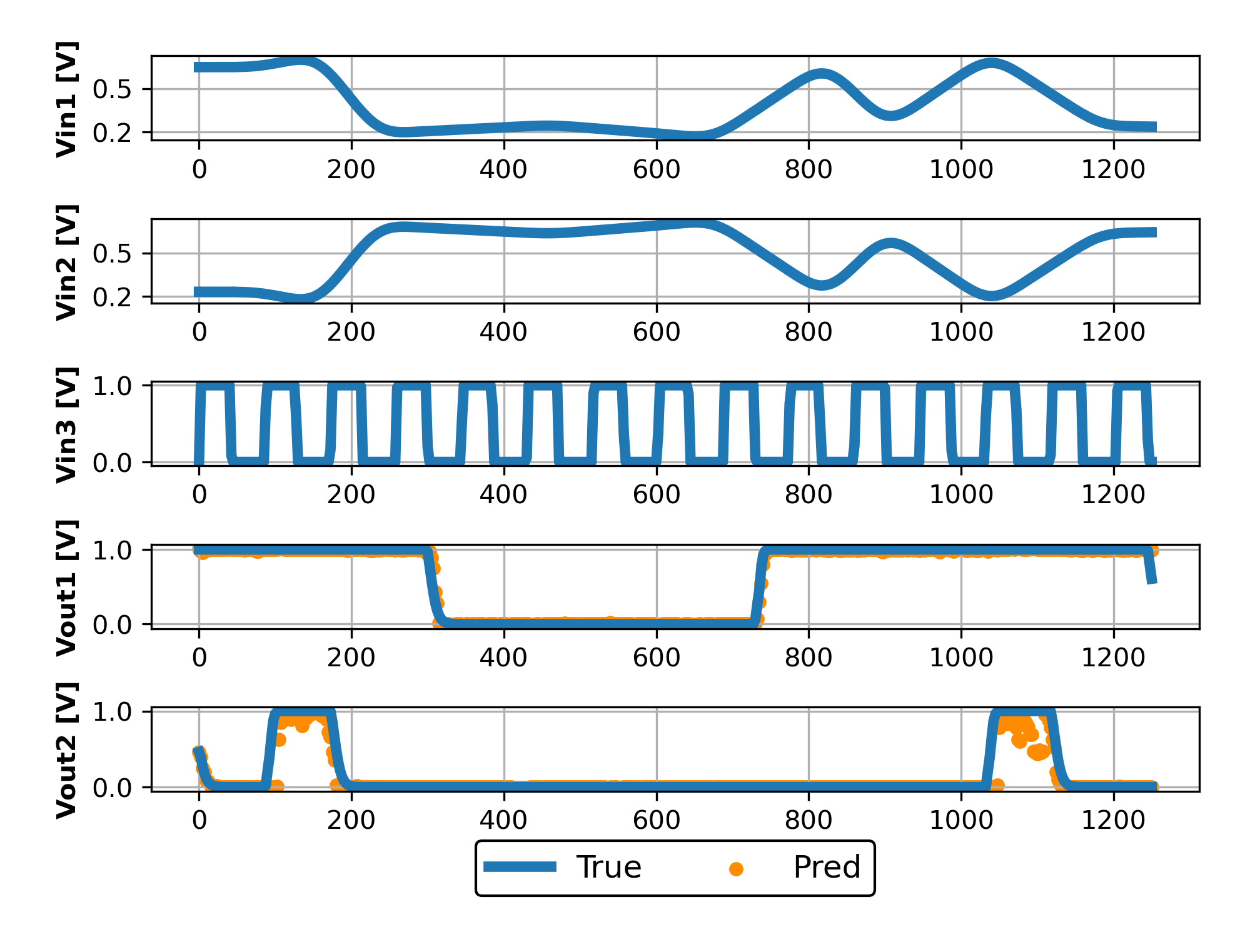}
         \caption{Crossformer + KAN Prediction}
     \end{subfigure}
     \hfill
     \caption{Fitting result comparison on one example data record. Crossformer + KAN prediction has a more stable prediction for output 1 (Vout1) as it captures well the stiffness and the plateau. For output 2 (Vout2), Crossformer + KAN has a more accurate prediction on stiff signal changes and it reduces the fluctuations shown in the CTRNN. }
     \label{fitting_compare}
\end{figure}

\begin{figure}
     \centering
     \begin{subfigure}[b]{0.41\textwidth}
         \centering
         \includegraphics[width=\textwidth]{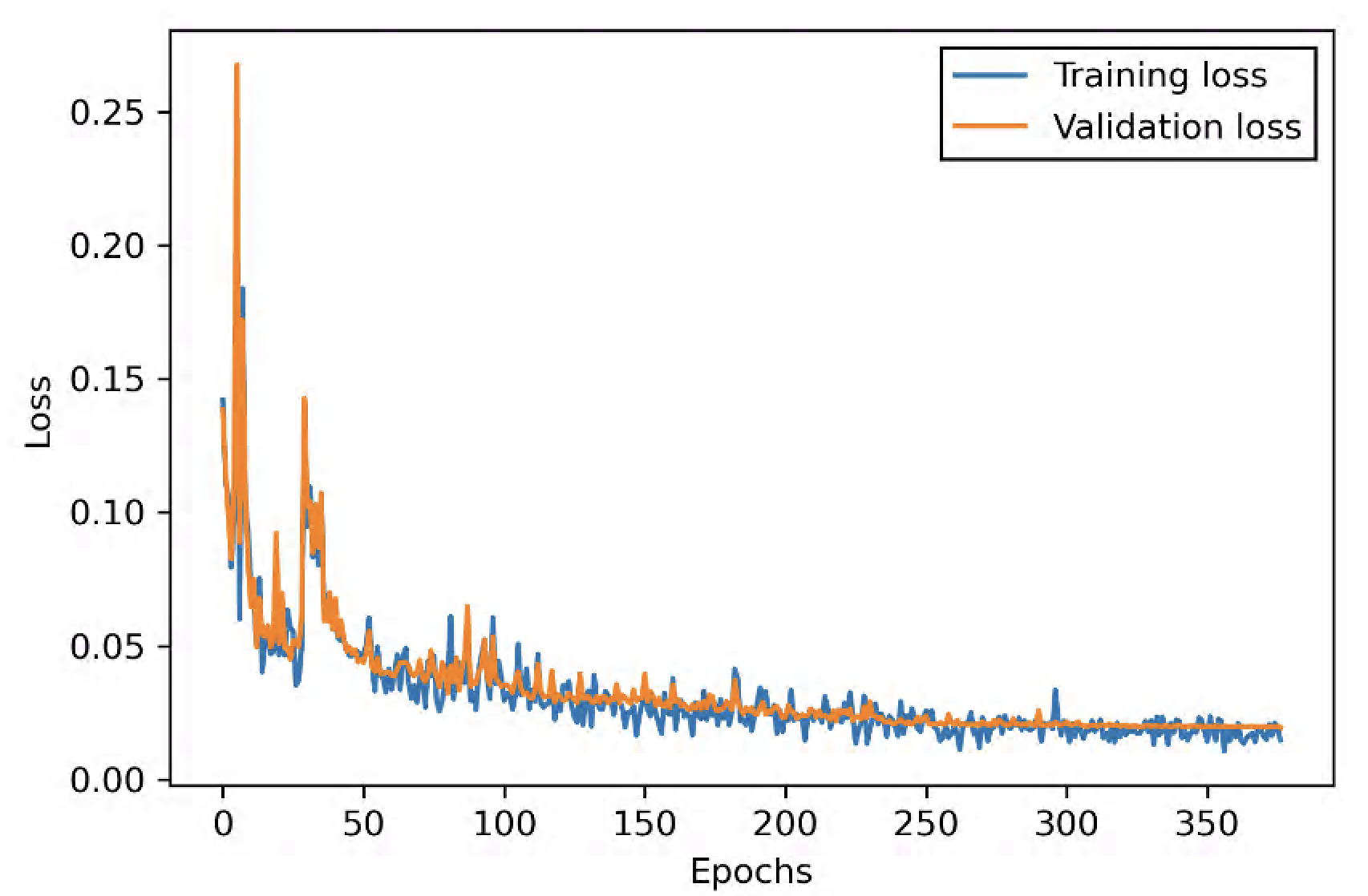}
         \caption{Loss curves of CTRNN training}
     \end{subfigure}
     \begin{subfigure}[b]{0.48\textwidth}
         \centering
         \includegraphics[width=\textwidth]{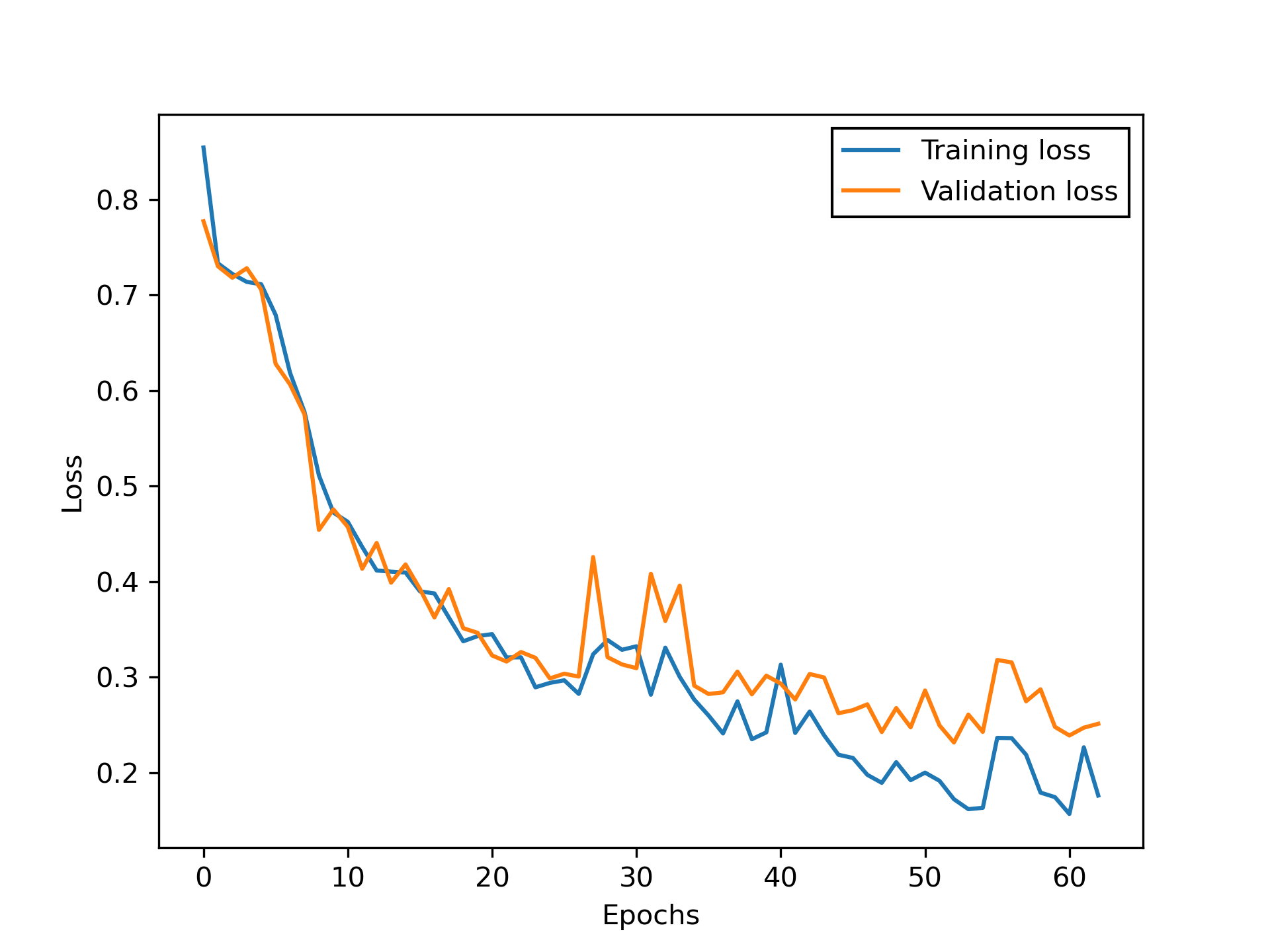}
         \caption{Loss curves of Crossformer training}
     \end{subfigure}
     \hfill
     \begin{subfigure}[b]{0.48\textwidth}
         \centering
         \includegraphics[width=\textwidth]{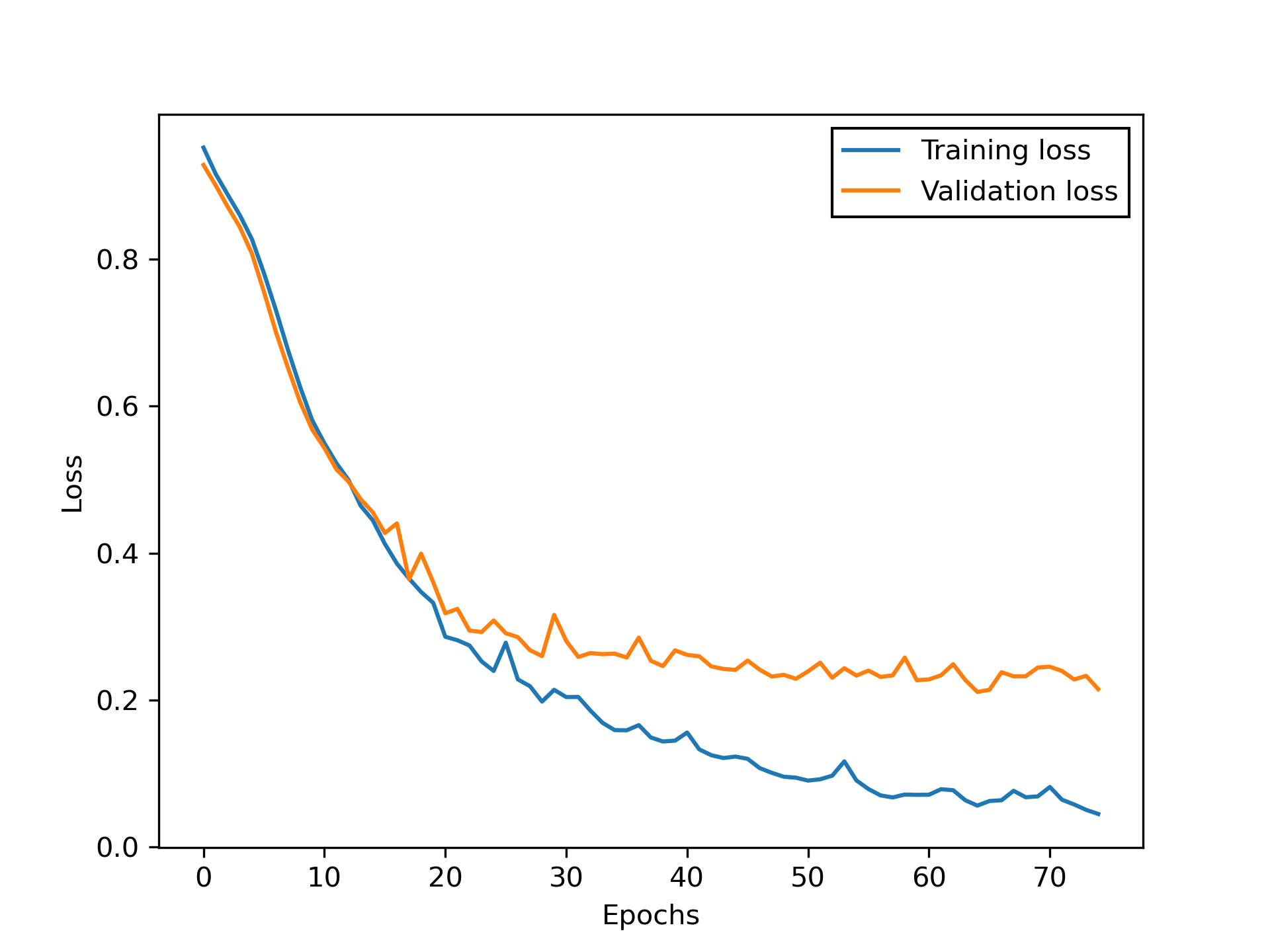}
         \caption{Loss curves of Crossformer + KAN training}
     \end{subfigure}
     \caption{Learning curves comparison for all models. CTRNN has requires more epochs to converge than Crossformer only and Crossformer + KAN. By adding KAN with increased neurons to Crossformer, the model becomes more powerful in learning the training data, leading to the minimum validation loss among all models.}
     \label{training_compare}
\end{figure}

From the convergence rate shown in Fig. \ref{training_compare}, we can see that our model can finish training within 65 epochs. However, CTRNN model requires more than 350 epochs to obtain the best model. This is another evidence that attention mechanism and KANs in our model helps the model capture the dependency of the data. In addition, since ODE solver cannot be accerlerated by GPU, the actual training time of CTRNN would be much longer compared to our model. Also, it is worth to note that CTRNN training lacks stability. Vanishing gradient problem happened a few times during our experiment so the actual time it requires to obtain a CTRNN model is much longer than that of our model.

\section{Discussion}
To better understand why our model provides a better fitting result, we can look into the individual model predictions made by CTRNN and our model (Fig. ~\ref{fitting_compare}). We can see from the plot that CTRNN prediction is more noisy compared to our model prediction when the ground truth signal stays at 0 or 1. This smoothness is likely due to the DSW Embedding since it uses signal segments instead of single points for output prediction. In addition to DSW embedding, KANs also help improve the result. Historically, the input and output relationship of the circuit can be accurately modeled using physics equations. Therefore, since KANs demonstrate excellent performance on other physics deep learning tasks, it is expected to work well for the circuit modeling problem.

\section{Conclusion}
In this work, we proposed a transformer-based circuit behavioral model that integrates KANs and Crossformer, achieving superior performance over the prior state-of-the-art CTRNN model for stiff circuit modeling. Under the same dataset, our approach delivers higher prediction accuracy while requiring less training time.

Looking forward, one promising direction is to incorporate more domain knowledge~\cite{wangwei1, hy1, wu1, weiman1, weiman2}, such as Fourier-based representations ~\cite{wu2} in the feature embedding stage, to further enhance model performance. Another potential avenue is to introduce adaptive refinement mechanisms during training and inference, inspired by recent modular self-improving and planning-based frameworks in machine learning~\cite{yang2026evotool, jasen1, jasen2, yang2026tooltree, jasen5}. 

Finally, we note that, unlike many other popular research areas \cite{hy2, wangwei2, weiman4, li4, li5}, there is currently no publicly accessible dataset specifically designed for stiff circuit modeling. Establishing standardized public datasets would greatly facilitate benchmarking and promote further research in this domain. Data augmentation \cite{li1, li2, li3} or active learning \cite{weiman3} technique might be used as a temporary solution for the lack of data.

\bibliographystyle{ACM-Reference-Format}
\bibliography{ref1}

\clearpage
\appendix


\begin{figure*}[!htbp]
  \centering
  \begin{minipage}{0.48\textwidth}
    \section{Appendix}
    \subsection{Hyperparameter tuning with KAN}
    \centering
    \begin{subfigure}[b]{\textwidth}
      \centering
      \includegraphics[width=0.9\textwidth]{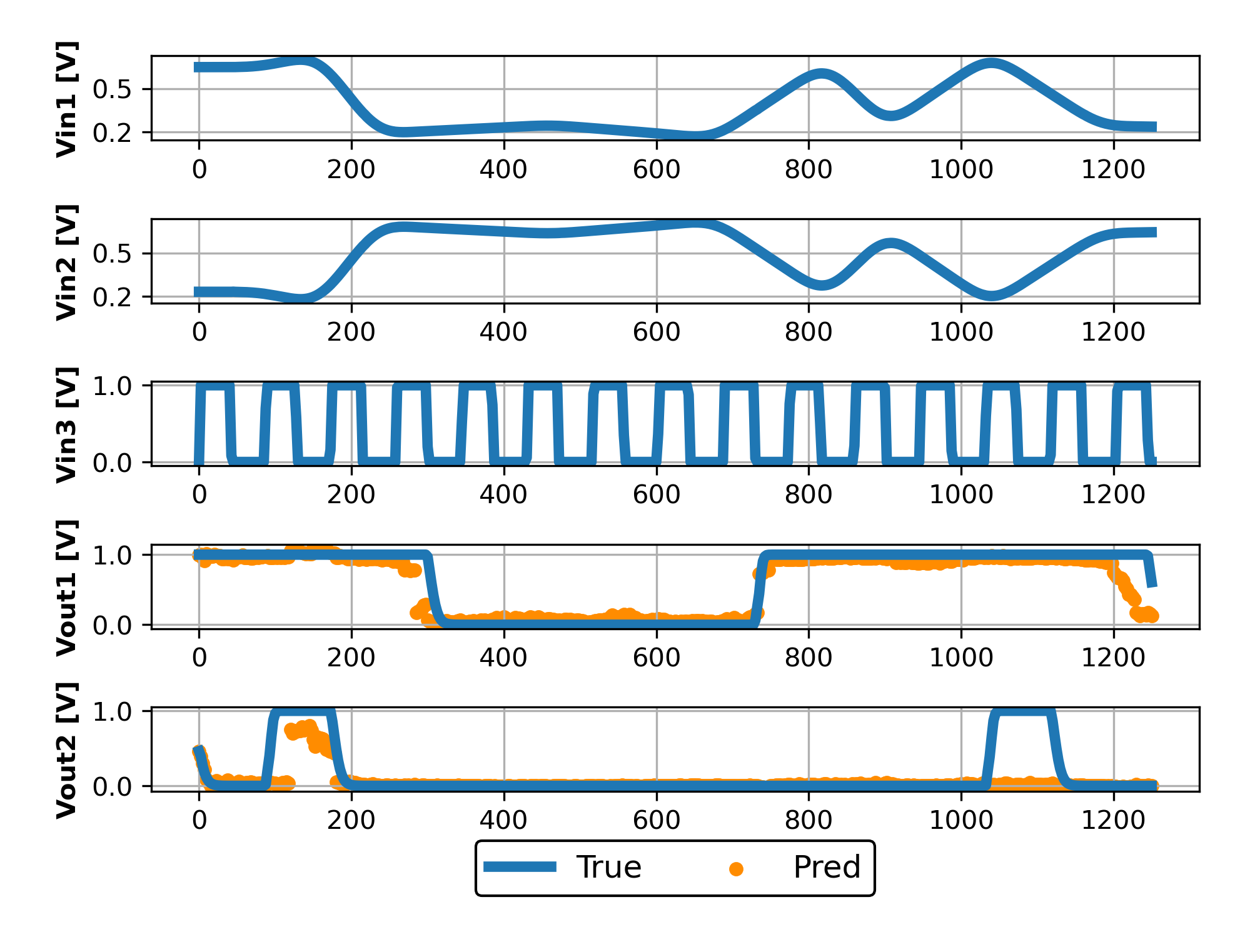}
      \caption{neurons=5, grid interval=5, k=3}
    \end{subfigure}
    
    \vspace{0.3cm}
    
    \begin{subfigure}[b]{\textwidth}
      \centering
      \includegraphics[width=0.9\textwidth]{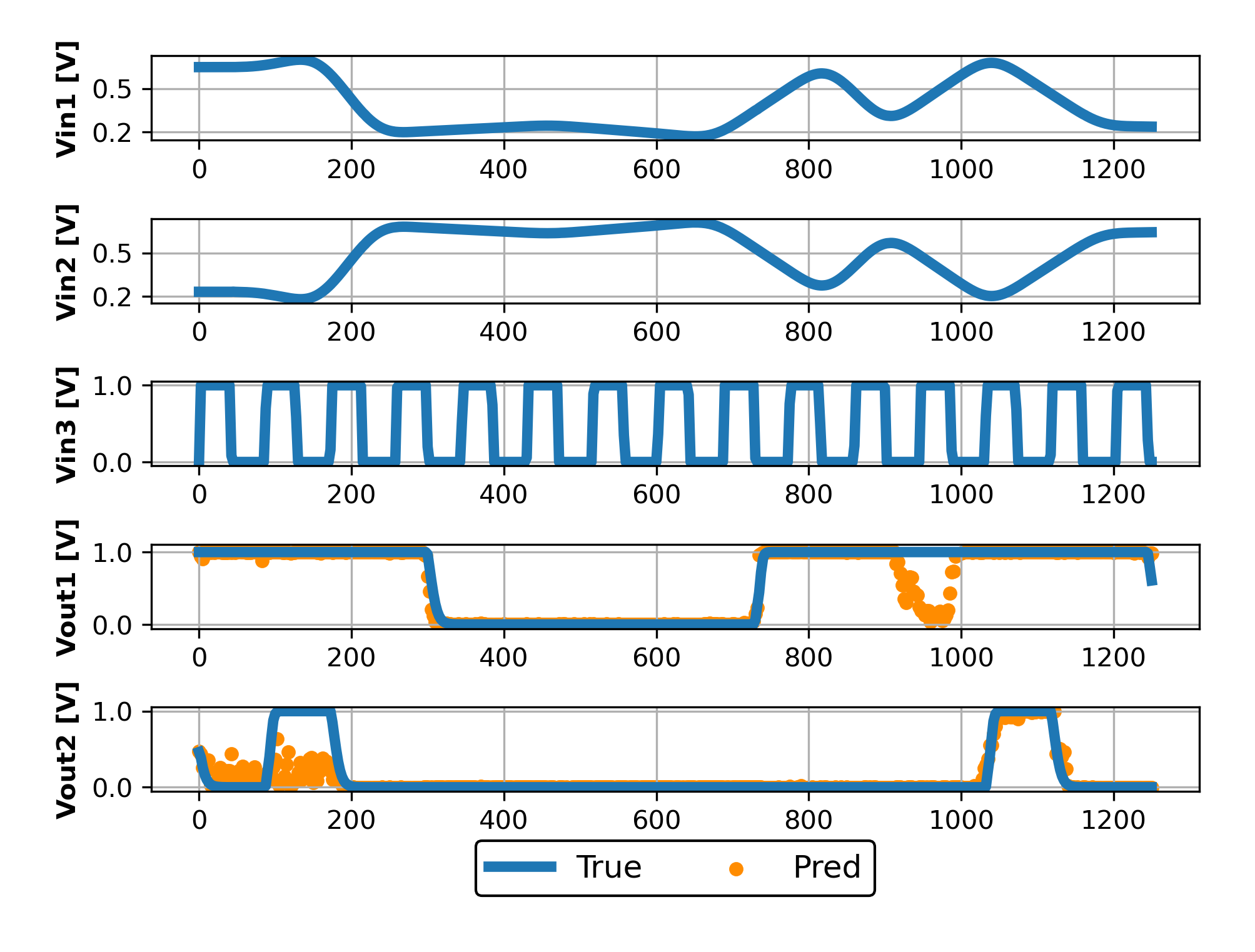}
      \caption{neurons=5, grid interval=50, k=3}
    \end{subfigure}
    
    \vspace{0.3cm}
    
    \begin{subfigure}[b]{\textwidth}
      \centering
      \includegraphics[width=0.9\textwidth]{figures/prediction_kan_n10_grid5_k3.png}
      \caption{neurons=10, grid interval=5, k=3}
    \end{subfigure}
    
    \caption{Prediction results}
    \label{kan_prediction}
  \end{minipage}
  \hfill
  \begin{minipage}{0.48\textwidth}
    \centering
    \begin{subfigure}[b]{\textwidth}
      \centering
      \includegraphics[width=0.9\textwidth]{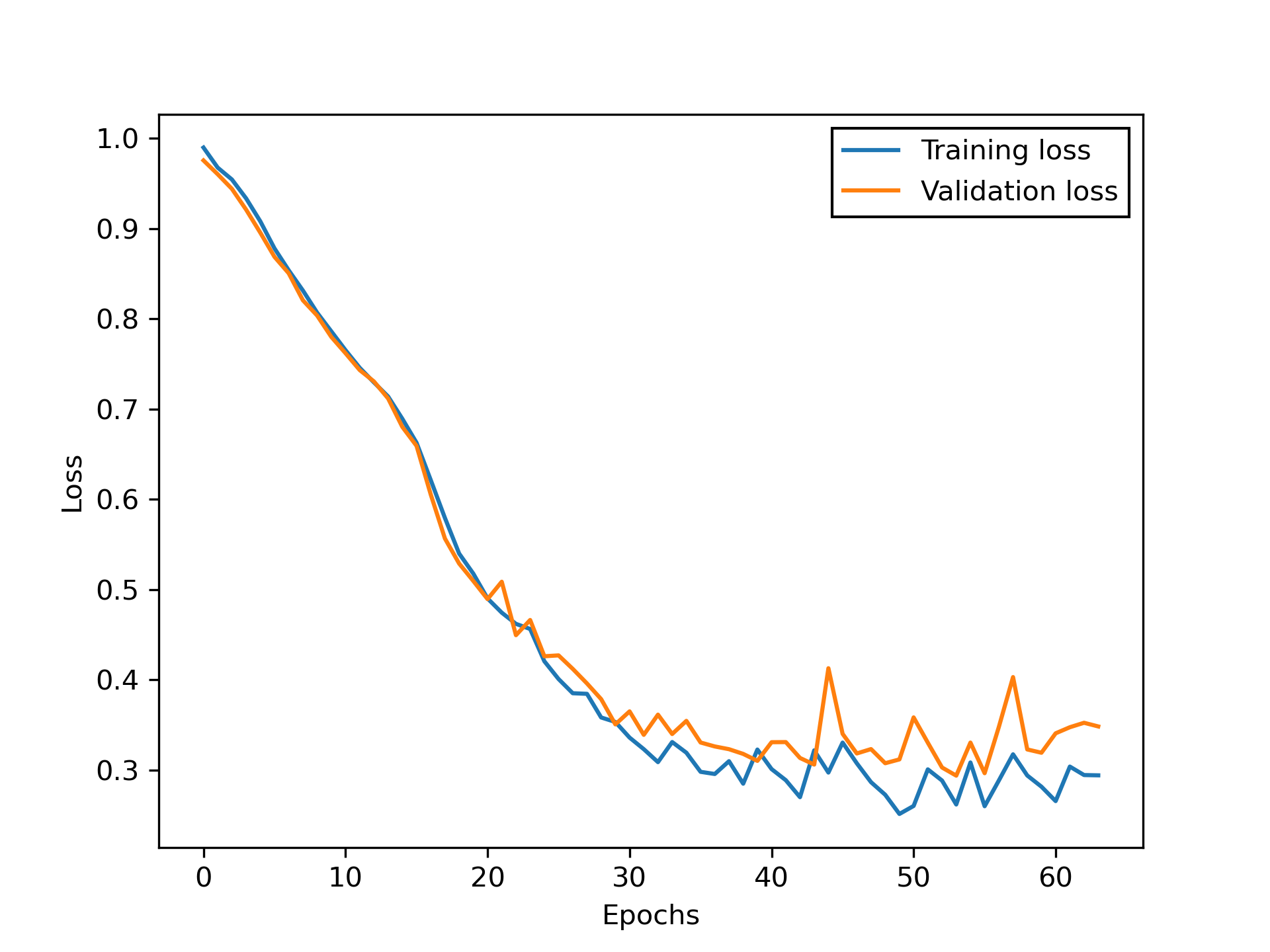}
      \caption{neurons=5, grid interval=5, k=3}
    \end{subfigure}
    
    \vspace{0.3cm}
    
    \begin{subfigure}[b]{\textwidth}
      \centering
      \includegraphics[width=0.9\textwidth]{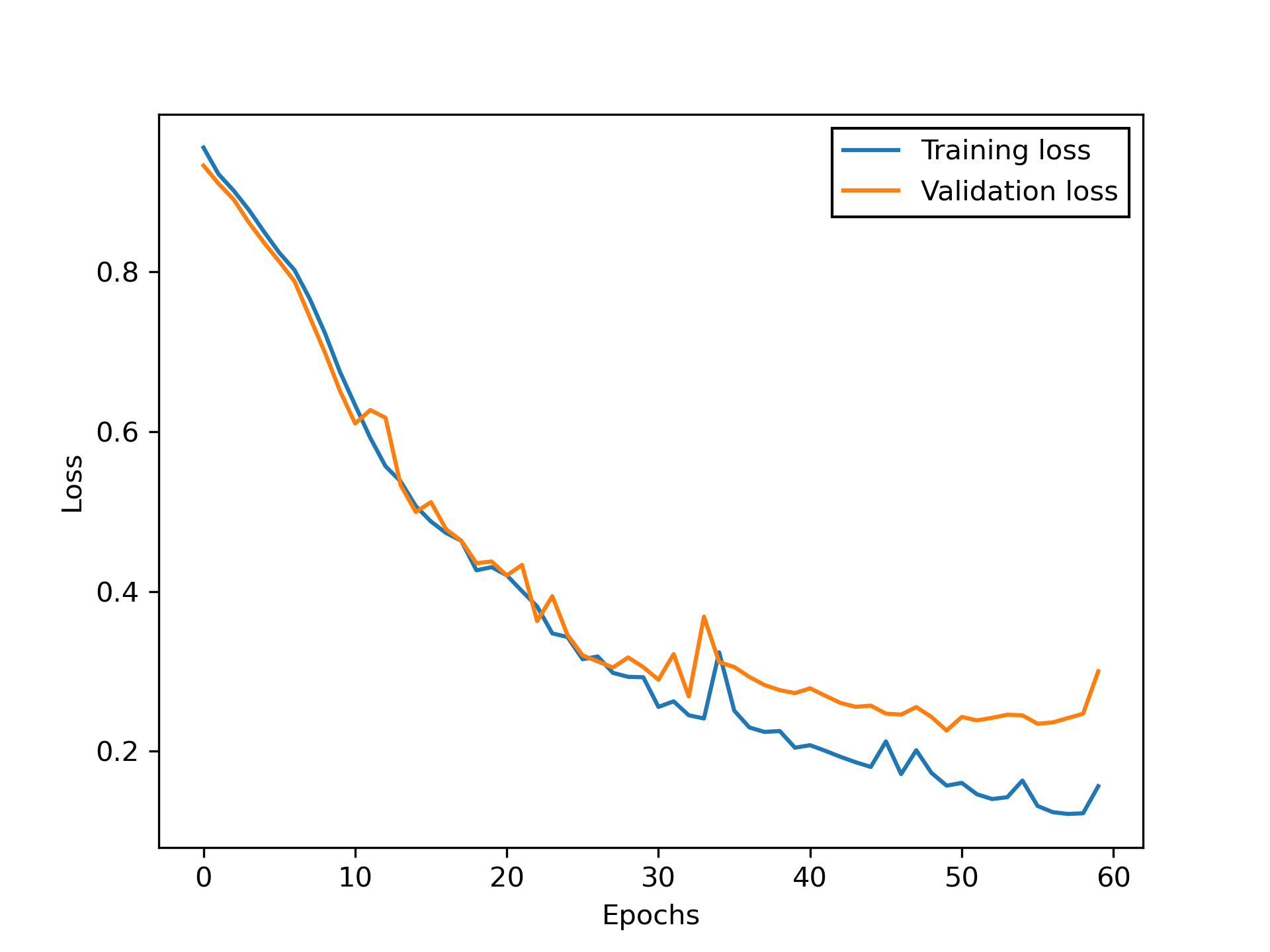}
      \caption{neurons=5, grid interval=50, k=3}
    \end{subfigure}
    
    \vspace{0.3cm}
    
    \begin{subfigure}[b]{\textwidth}
      \centering
      \includegraphics[width=0.9\textwidth]{figures/loss_kan_n10_grid5_k3.png}
      \caption{neurons=10, grid interval=5, k=3}
    \end{subfigure}
    
    \caption{Learning curves}
    \label{kan_loss}
  \end{minipage}
\end{figure*}

\end{document}